\documentclass[aps,twocolumn,english,aip,superscriptaddress,amssymb, showpacs,preprintnumbers,amsmath,amssymb,superscriptaddress]{revtex4-1}

\usepackage{graphicx}
\usepackage{dcolumn}
\usepackage{bm}
\newcommand{\nc}{\newcommand}
\nc{\be}{\begin{equation}}
\nc{\ee}{\end{equation}}
\nc{\bea}{\begin{eqnarray}}
\nc{\eea}{\end{eqnarray}}
\nc{\bean}{\begin{eqnarray*}}
\nc{\eean}{\end{eqnarray*}}
\nc{\mb}{\mbox}
\nc{\rnc}{\renewcommand}
\nc{\vk}{\mb{\bf k}}
\nc{\vp}{\mb{\bf p}}
\nc{\vn}{\mb{\bf n}}
\nc{\vq}{\mb{\bf q}}
\nc{\rr}{\mb{\bf r}}
\nc{\vz}{\hat {\mb{\bf z}}}
\nc{\vj}{\mb{\boldmath$j$}}
\nc{\vg}{\mb{\boldmath$g$}}
\nc{\x}{\mb{\boldmath$x$}}
\nc{\A}{\mb{\boldmath$A$}}
\nc{\va}{\mb{\boldmath$a$}}
\nc{\vs}{\mb{\boldmath$\sigma$}}
\nc{\vpi}{\mb{\boldmath$\pi$}}
\nc{\nab}{\nabla}
\nc{\X}{\sf x}

\begin{document}
\title{Binding a Hopfion in Chiral Magnet Nanodisk}

\author{Yizhou Liu}
\thanks{yliu062@ucr.edu\\
Present Address: Beijing National Laboratory for Condensed Matter Physics, Institute of Physics, Chinese Academy of Sciences, Beijing 100190, China}
\affiliation{Department of Electrical and Computer Engineering, University of California, Riverside, California 92521, USA}

\author{Roger K. Lake}
\thanks{rlake@ece.ucr.edu}
\affiliation{Department of Electrical and Computer Engineering, University of California,
Riverside, California 92521, USA}

\author{Jiadong Zang}
\thanks{Jiadong.Zang@unh.edu}
\affiliation{Department of Physics and Materials Science Program, University of New Hampshire, Durham, New Hampshire 03824, USA}

\begin{abstract}

Hopfions are three-dimensional (3D) topological textures characterized by the integer Hopf invariant $Q_H$. 
Here, we present 
the realization of a zero--field, stable hopfion spin texture 
in a magnetic system consisting of a chiral magnet nanodisk sandwiched by two films with perpendicular magnetic anisotropy. 
The preimages of the spin texture and numerical calculations of $Q_H$ show that the hopfion has $Q_H=1$.
Furthermore, another non-trivial state that includes a monopole--antimonopole pair (MAP) is also stabilized in this system.
By applying an external magnetic field, hopfion and MAP states with the same polarization can be switched between each other.
The topological transition between the hopfion and the MAP state involves a creation (annihilation) of the MAP and twist of the preimages.
Our work paves the way to study non-trivial 3D topological spin textures and stimulates more investigations in the field of 3D spintronics.

\end{abstract}


\maketitle

A topological soliton carries an integer topological index that cannot be changed by a continuous deformation~\cite{manton_topological_2004}. 
A celebrated example is the skyrmion, 
a two-dimensional (2D) topological soliton originated from the Skyrme model~\cite{skyrme_unified_1962}, 
which can be  characterized by the skyrmion number (or winding number)~\cite{Rajaraman_1987}.
The addition of a third spatial dimension brings more diverse and complicated topological solitons, such as rings, links 
and knots~\cite{volovik_1977, battye_knots_1998, battye_solitons_1999}.
Some of these three-dimensional (3D) topological solitons are ``hopfions'', 
since they can be classified by the Hopf invariant ($Q_H$) ~\cite{hopf_uber_1931}, 
a topological index of the homotopy group $\Pi_3(S^3)$ that can be interpreted as the linking number~\cite{link_hopf}.
Due to their complex structures and models, 
the detailed study of the hopfion was properly established not long ago 
in terms of toroidal coordinates\cite{faddeev_stable_1997,faddeev_toroidal_1997}.
Hopfions have been observed in a variety of physical systems including fluids, optics, liquid crystals, Bose-Einstein condensates, etc.~\cite{kleckner_creation_2013, kleckner_how_2016, dennis_isolated_2010, kawaguchi_knots_2008, ackerman_self_assembly_2015, ackerman_static_2017, hall_tying_2016}
But their observation in magnetic materials remains elusive.

In magnetic systems, topological solitons in one dimension and two 
dimensions such as domain walls and vortices have been extensively studied over the past few decades.
Much of the recent attention is attracted by the magnetic skyrmions residing in magnetic materials 
with the antisymmetric Dzyaloshinskii-Moriya interaction (DMI)~\cite{dzyaloshinsky_thermodynamic_1958, moriya_anisotropic_1960, rosler_spontaneous_2006}.
The spins of a magnetic skyrmion wind around the unit sphere once, which results in the unit winding number of a skyrmion.
Skyrmions are proposed to be promising candidate for spintronic 
applications due to their prominent features such as the nanoscale size and 
low driving current density~\cite{fert_skyrmions_2013, jonietz_spin_2010}.

Although numerous studies have been made on the low-dimensional topological solitons, 
3D topological solitons like hopfions have still not been well explored in nanomagnetism.
Understanding the static and dynamical properties of these 3D topological 
solitons are not only of fundamental interest, but may also enable future applications.
Only a few theoretical proposals predict the existence of hopfions in ferromagnets, 
but only in the dynamical regime~\cite{cooper_propagating_1999,sutcliffe_vortex_2007,borisov_stationary_2008}.
It has been recently proposed that 
a higher order exchange interaction and an external magnetic field
will stabilize a metastable hopfion in a frustrated magnet~\cite{sutcliffe_skyrmion_2017}, but how to create such metastable state is not clear.
%
%

In this Letter, we show that a $Q_H = 1$ hopfion can be enabled 
in a chiral magnet nanodisk in the absence of external magnetic fields. 
The nanodisk is sandwiched by two magnetic 
layers with perpendicular magnetic anisotropy (PMA) to nucleate the hopfion in between. 
The hopfion is identified by both the preimages and the numerical calculations of $Q_H$.
Associated with the hopfion, 
another non-trivial state that includes a monopole-antimonopole pair (MAP) 
is also stabilized at zero fields in this structure.
Furthermore, the hopfion can be switched into a MAP state with the same polarization by an applied magnetic field, and vice versa.
The topological transition between the hopfion state and the MAP state 
involves the creation (annihilation) of the monopole-antimonopole pair and a twist of the preimages.

\begin{figure}
\begin{center}
\includegraphics[width=2.8in,height=4.1in]{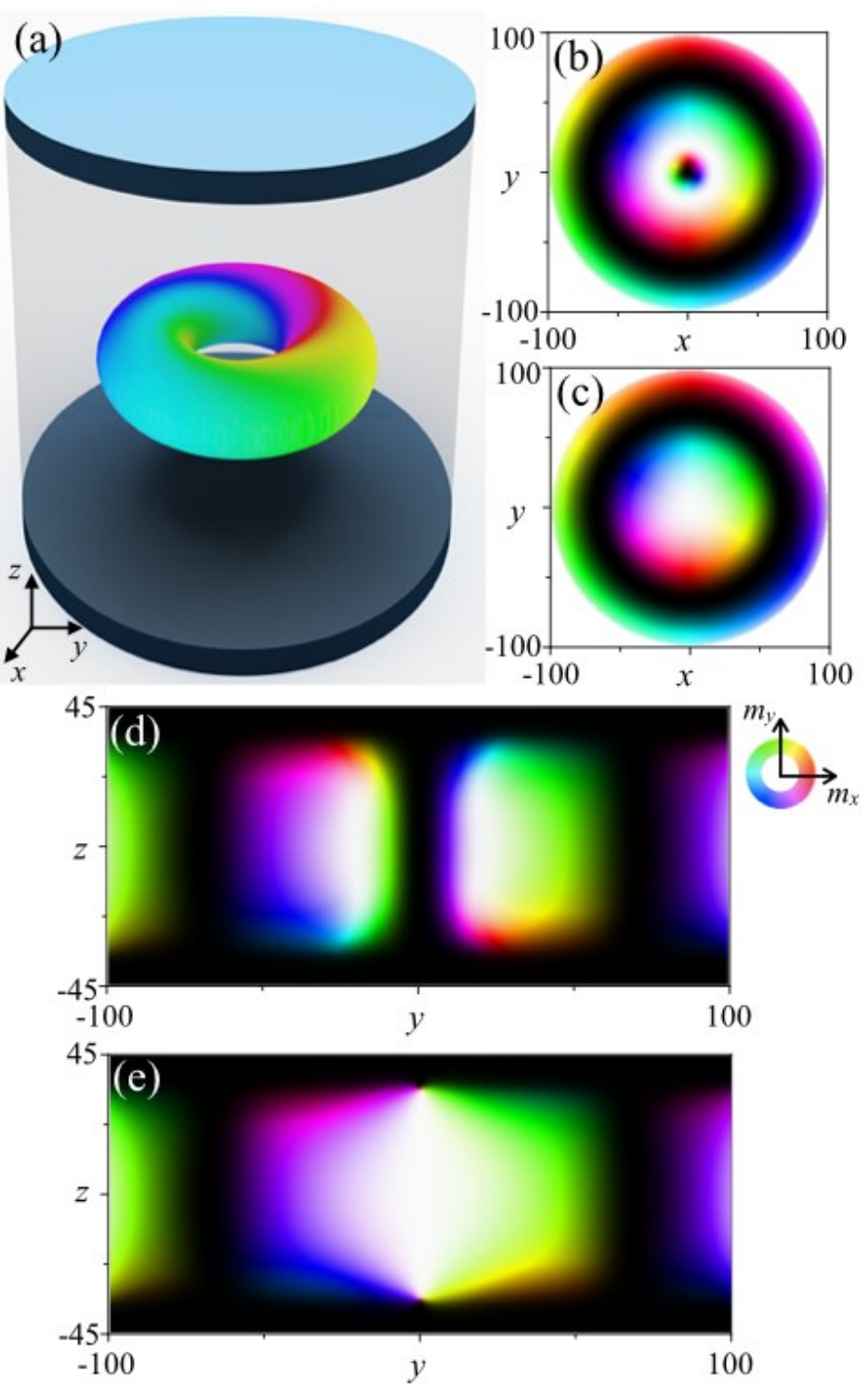}
\caption{(a) Schematic of the proposed structure. 
The thin disks at the top and bottom represent the magnetic films with PMA. 
The transparent region in the middle is the chiral magnet nanodisk. 
The color ring at the center represents the set of preimages with $s_z = 0$ of a $Q_H = 1$ hopfion. 
(b), (c) The cross-sectional spin textures in the x-y plane (z=0) for the hopfion (b) and MAP (c). 
(d), (e) The cross-sectional spin textures in the y-z plane (x=0) for the hopfion (d) and MAP (e). 
In the color scheme, black indicates $s_z=-1$ and white indicates $s_z=1$. 
The color wheel is for $s_z=0$.}
\label{fig:schematic}
\end{center}
\end{figure}

We  consider a chiral magnet nanodisk with radius 100 nm and thickness 70 nm sandwiched by two PMA magnetic thin layers with 10 nm thickness, as shown in Fig.~\ref{fig:schematic}(a). 
An isotropic bulk type DMI is employed to model the chiral magnet.
The Hamiltonian of this system is given by
\begin{equation}
{\cal H} = \int  d r^3 [ -A (\nabla \textbf{s} )^2 - (1-{\mbox{p}})D \textbf{s} \cdot ( \nabla \times 
\textbf{s}) - {\mbox{p}} K_u  (s_z)^2 + E_{\rm d}
],
\label{eq:energy}
\end{equation}
where $A$ and $D$ are the exchange and DMI constant, respectively, $K_u$ is the PMA constant, and $\mbox{p}$ is 0 in the chiral magnet nanodisk and 1 in two PMA layers.
$E_{\rm d}$ is the magnetic dipole-dipole interaction (DDI).
It depends on the exact shape of the system. 
When the system size goes down to nanoscale, the DDI becomes important in determining the corresponding spin textures.
For example, the DDI favors the stabilization of magnetic skyrmion at zero-field in confined geometries.
It leads to the formation of the so-called target skyrmion, which has been theoretically proposed and recently experimentally observed in magnetic nanodisks without any external fields~\cite{du_target, leonov_target_skyrmions_2014, beg_ground_2015, zheng_direct_2017}.
Thus, the effect of DDI is essential in confined systems and cannot be ignored.

We minimize the Hamiltonian (\ref{eq:energy}) in the nanodisk structure with different initial states (for details of the simulation methods and parameters, see Supplemental Materials~\footnote{See Supplemental Materials, for details of the simulations, Hopf invariant calculations and movies for the topological transition. Micromagnetic simulations were performed using Mumax3~\cite{mumax_2014}, Fidimag~\cite{fidimag_2016} and an in-house micromagnetic simulation package.}).
After minimizing the energy, we find two stable non-trivial states at zero-field, 
the hopfion state and the MAP state.
The hopfion state includes a $Q_H = 1$ hopfion, and the MAP state includes a monopole-antimonopole pair. 
%

%
To present the detailed spin textures of the hopfion and MAP, 
cross-sections of both states are plotted in Fig.~\ref{fig:schematic}.
For the hopfion, the cross-section in the x-y plane (z=0) shown in Fig. \ref{fig:schematic}(b), 
has a skyrmion at the center surrounded by two concentric spin helical rings.
This is typically a target skyrmion configuration recently observed in an FeGe nanodisk \cite{zheng_direct_2017}.
A conventional skyrmion is wrapped by a concentric helical ring.
From the center to periphery, spin rotates by an angle of $2\pi$ instead of $\pi$ in typical skyrmions.
The outmost spin helical ring is not part of the hopfion but an edge state induced 
by the DDI from the circular shape and the DMI of the chiral magnet.
The lateral cross-section in the y-z plane (x=0) shown in Fig. \ref{fig:schematic}(d) 
includes a skyrmion--antiskyrmion pair.
The cross-section taken at any plane containing the $z$-axis always contains a 
skyrmion--antiskyrmion pair.
This is a result of the hopfion spin texture that consists of a 
$2\pi$ twisted skyrmion tube with its two ends 
glued together as shown in Fig.~\ref{fig:schematic}(a).
For the MAP state, the cross-section in the x-y plane (z=0) shown in Fig. \ref{fig:schematic}(c) 
is a typical skyrmion.
%
The cross-section in the y-z plane (x=0) shown in Fig. \ref{fig:schematic}(e), 
has only one spin up region, in contrast to the
skyrmion--antiskyrmion pair of the hopfion. 
Instead, a monopole (antimonopole) is formed near the top (bottom) surface.
This originates from the restricted spin polarization of the PMA layers on the top and bottom.

\begin{figure}
\begin{center}
\includegraphics[width=3.4in,height=2.07in]{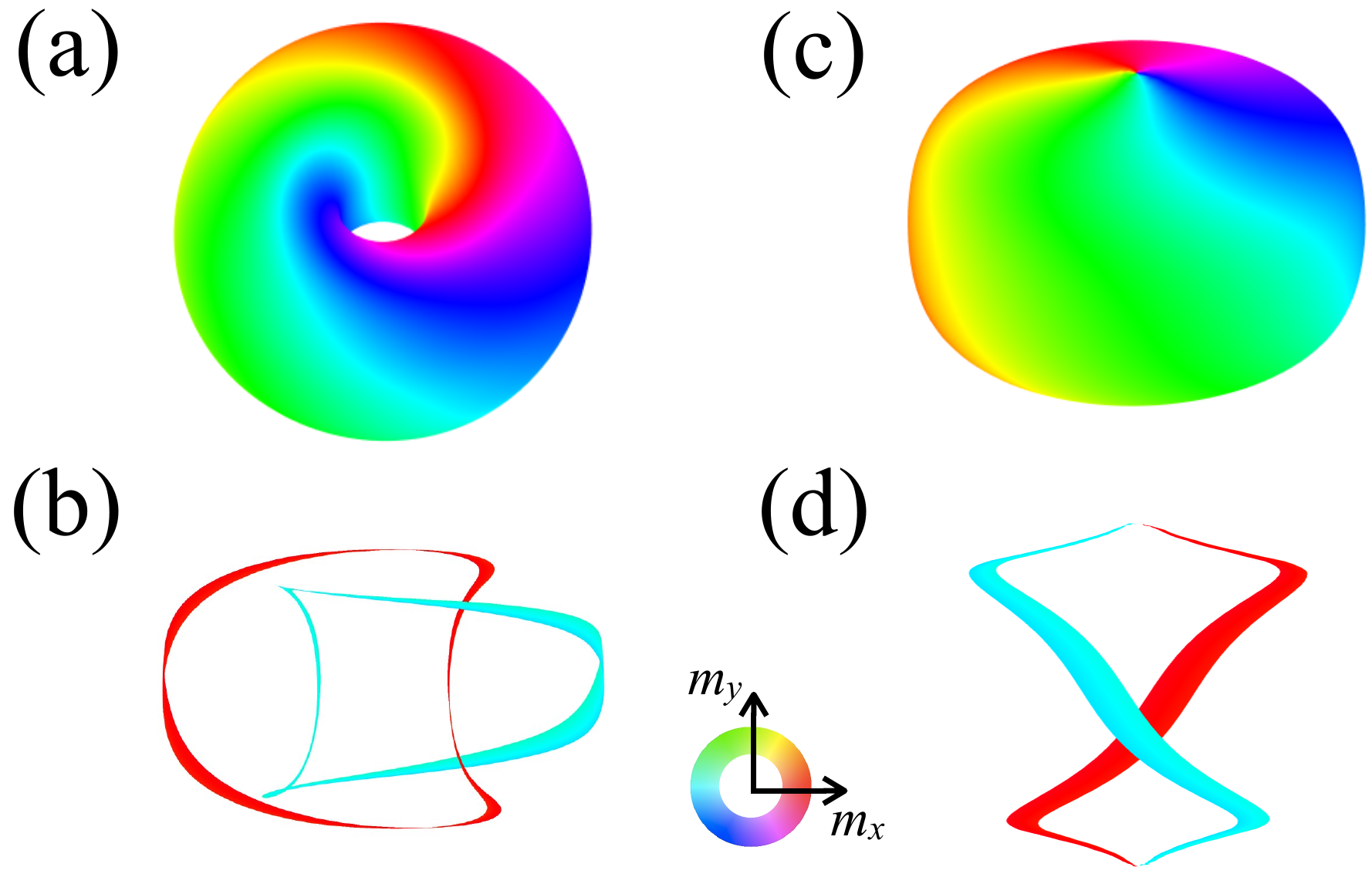}
\caption{(a), (c) The set of preimages with $s_z = 0$ for the hopfion and MAP, respectively. 
(b), (d) The preimages of $\bf s=$(-1,0,0) (cyan) and $\bf s=$(1,0,0) (red) for the hopfion (b) and MAP (d). 
In the color scheme, black indicates $s_z=-1$ and white indicates $s_z=1$. 
The color wheel is for $s_z=0$.}
\label{fig:preimage}
\end{center}
\end{figure}

To further visualize and understand the spin configurations of the hopfion and MAP in 3D, 
we plot their preimages using Spirit~\cite{spirit}.
A preimage is the region in 3D real space that contains spins with the same orientations. 
It is a Hopf map of a point on the $S^2$ unit sphere to 3D space.
We first plot the set of preimages of all spins with $s_z = 0$ for the hopfion (Fig~\ref{fig:preimage}(a)) 
and MAP (Fig.~\ref{fig:preimage}(c)), 
which corresponds to a Hopf map from the equator of the $S^2$ unit sphere to the 3D space.
Two preimages are topologically distinct as characterized by different genus $g$, i.e., the number of holes.
The preimage of the hopfion forms a torus with $g=1$, whereas the preimage of the MAP is a trivial surface with $g=0$, which satisfies the Poincar\'e-Hopf theorem~\cite{Poincare_Hopf}. 

The Hopf invariant, also called the linking number, counts the number of links between 
two arbitrary closed-loop preimages.
Therefore, preimages of two arbitrary spins must form closed loops that are linked with each other.
These features can be identified by the preimages of $\textbf{s}=(1,0,0)$ and $\textbf{s}=(-1,0,0)$ 
for the hopfion (Fig.~\ref{fig:preimage}(b)) and MAP (Fig.~\ref{fig:preimage}(d)).
For the hopfion, two closed-loop preimages are formed and linked with each other once.
$Q_H = 1$ in this case, and the
topology of the hopfion state in this system is confirmed.
In contrast, the MAP does not have closed-loop preimages and thus no links.
Monopole and antimonopole are source and drain of all preimages.
The two MAP preimages of $\textbf{s}=(1,0,0)$ and $\textbf{s}=(-1,0,0)$
join at the monopole and antimonopole indicating their singular natures.
The MAP is considered a defect state, while the hopfion is a smooth spin texture with no singularity.
These preimages successfully reflect the topological natures of the two states.

\begin{figure}
\begin{center}
\includegraphics[width=2.7in,height=3.8in]{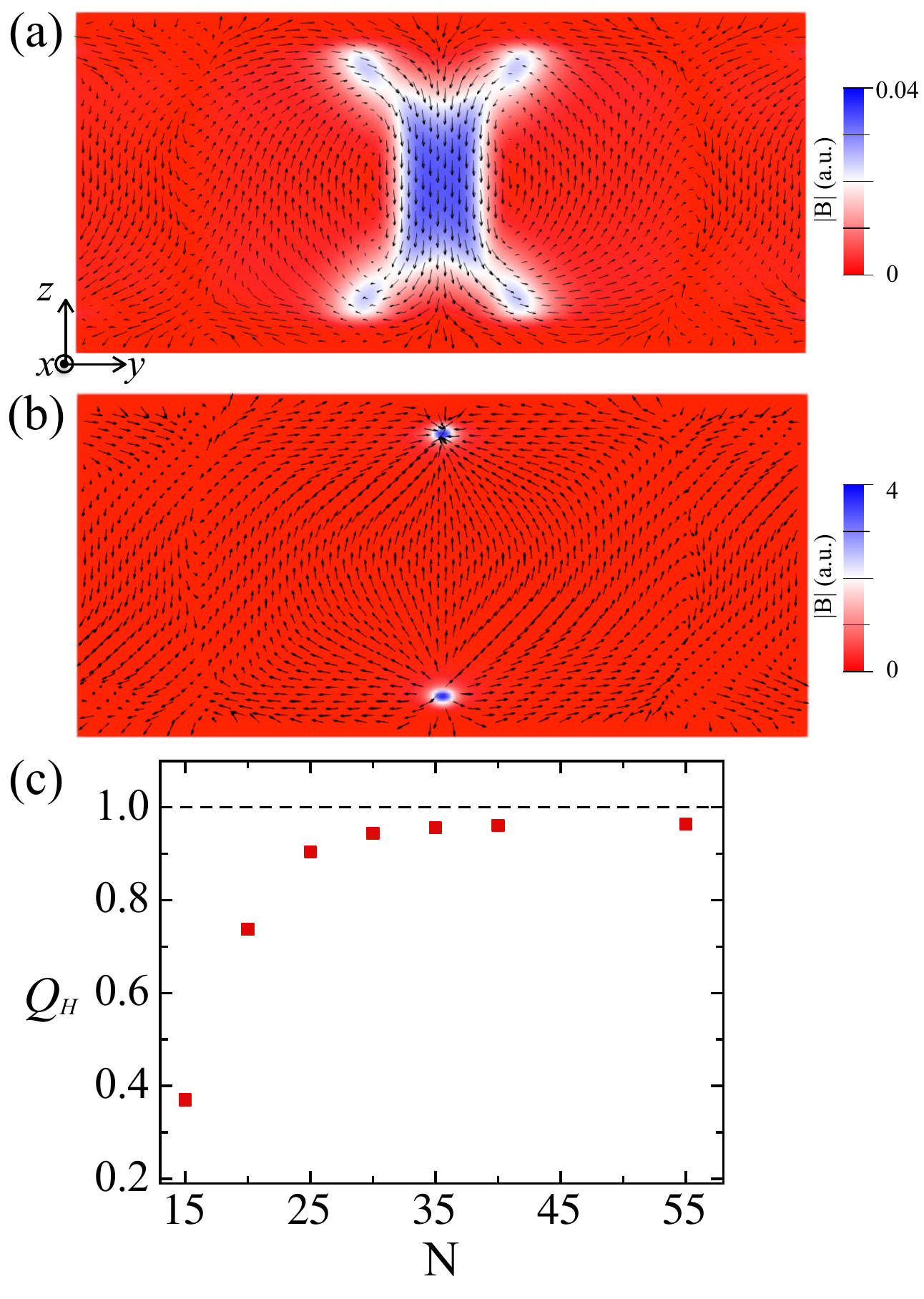}
\caption{(a), (b) The emergent magnetic field ${\bf B}$ in the y-z plane (x=0) for hopfion (a) and MAP (b). 
(c) Numerical calculations of the Hopf invariant $Q_H$ for different meshes. 
The total number of grid points $N_{tot} = N^3$. 
}
\label{fig:charge}
\end{center}
\end{figure}

Other than the linking number of preimages, 
topology of the hopfion can also be confirmed by directly calculating the Hopf invariant.
The integral form of the Hopf invariant in real space can be expressed as~\cite{whitehead_expression_1947,wilczek_linking_1983}
\begin{equation}
Q_H = - \int {\bf B} \cdot {\bf A} d\bf{r}, 
\label{eq:hopf_charge}
\end{equation}
where $B_i = \frac{1}{8\pi} \epsilon_{ijk} {\bf n} \cdot (\nabla_j {\bf n} \times {\nabla_k} {\bf n})$ 
is the emergent magnetic field associated with the spin textures, 
and $\bf A$ is any vector potential that satisfies the magnetostatic equation ${\bf \nabla \times A} = \bf B$.
The Hopf number is invariant under a gauge transformation ${\bf A}\rightarrow{\bf A}+\nabla\chi$ 
only when the emergent field $\bf B$ is free of singularities, {\it i.e.}, $\nabla\cdot{\bf B}=0$.
%
%
Cross-sections in the y-z plane of
the emergent magnetic fields ${\bf B}$ of the hopfion and MAP states are shown in Fig.~\ref{fig:charge}.
The emergent ${\bf B}$ field 
of the hopfion shown in Fig.~\ref{fig:charge}(a) 
flows smoothly 
and streams intensively near the center of the nanodisk. 
In contrast, the emergent ${\bf B}$ field of the MAP shown in Fig.~\ref{fig:charge}(b) 
clearly presents two magnetic monopoles with opposite charges near the top and bottom surface. 
%
The Hopf invariant is thus ill--defined for the MAP state, and it is well defined for the hopfion texture.
%
%

A gauge field solution ${\bf A}$ must also be solved 
in order to directly calculate $Q_H$ in real space.
%
To this end, we solve for the vector potential ${\bf A}$ in momentum space with the Coulomb gauge ${\bf k \cdot A}=0$, and then compute $Q_H$ in momentum space~\cite{moore_topological_2008}.
To carry out the numerical integral, discrete grids in the momentum space are employed.
As shown in Fig.~\ref{fig:charge}(c), as the grid number ($N_{tot}$) increases, $Q_H$ rapidly converges to 1.
We thus obtain a Hopf invariant of $Q_H=0.96$ for the hopfion spin texture under investigation. 
Here $Q_H$ is slightly deviated from an integer due to the finite size and open boundary condition. 
The manifold is not compact, as indicated by the edge state around the disk boundary. 
Nevertheless, the Hopf invariant is close to $1$, and the topological nature of the hopfion is further confirmed.
%

\begin{figure}
\begin{center}
\includegraphics[width=3in,height=5.04in]{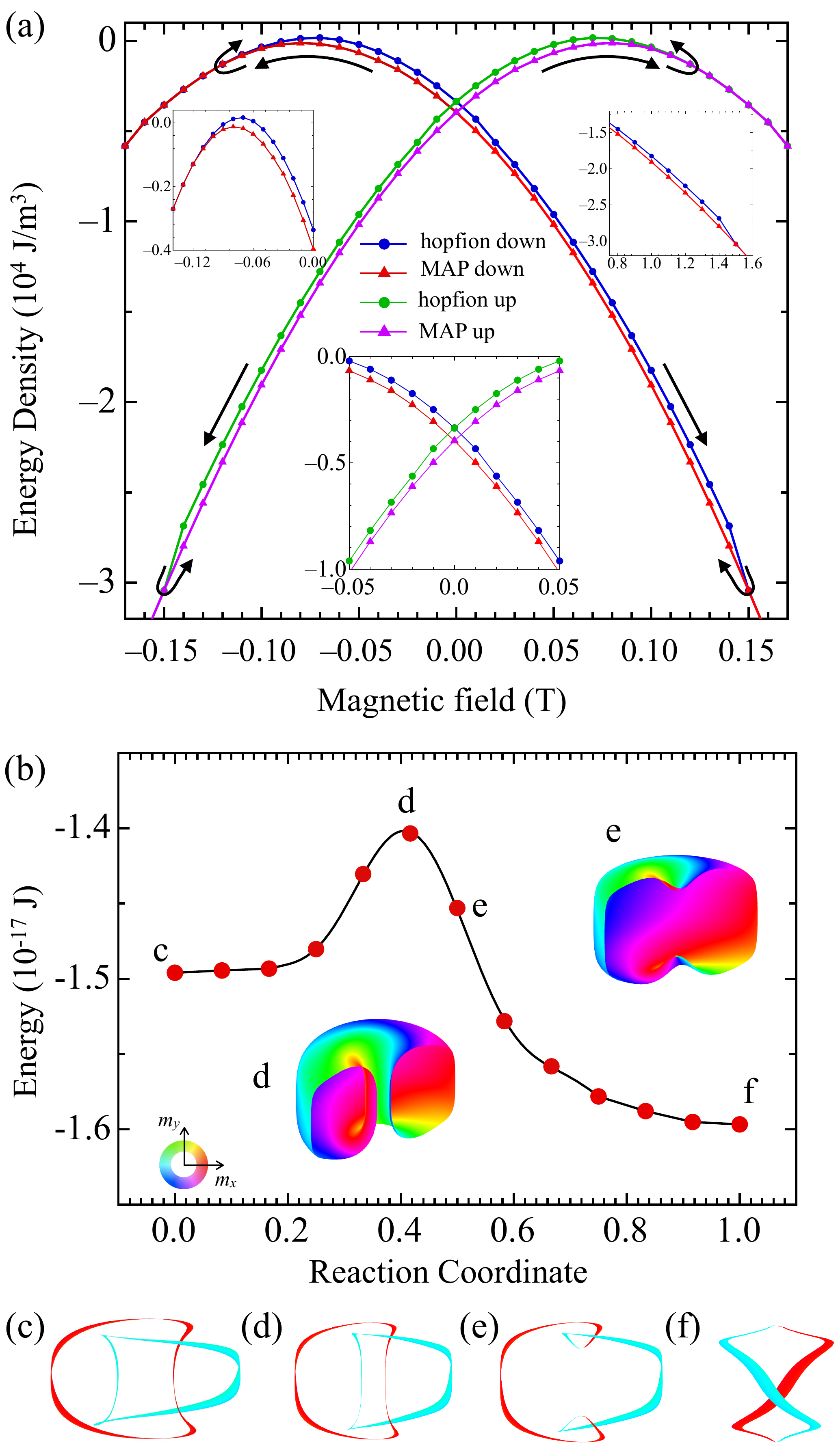}
\caption{(a) Energy density plot of the hopfion and MAP state as a function of external magnetic field. Black arrows indicate the field sweeping directions and the switching events.
Insets show the enlarged details of the plot.
(b) Minimal energy path between the hopfion and MAP state. 
Points c and f represent the hopfion and the MAP, respectively. 
The hopfion is nearly annihilated at
saddle point d, and the MAP is created at e.
Insets show 
the half-plane view preimages of $s_z=0$ for spin textures at d and e.
(c)-(f) The preimages of $\bf s=$(-1,0,0) (cyan) and $\bf s=$(1,0,0) (red) corresponding to points c--f in (a). 
}
\label{fig:mep}
\end{center}
\end{figure}

At zero external magnetic field, two states with opposite spins share the same energy.
Therefore, stable hopfion and MAP states each have two polarizations, 
i.e. spin points up or down at their cores.
As shown in Fig.~\ref{fig:mep}(a), the MAP state has lower energy than the hopfion state at zero magnetic field.
But they can be switched between each other by sweeping an external magnetic field.
When applying a magnetic field in the same (opposite) direction with the MAP (hopfion) polarization, the MAP (hopfion) can be switched into a hopfion (MAP) with the same polarization.
Thus, despite the MAP state having lower energy, the hopfion state can still be realized by using an applied field.
In Fig.~\ref{fig:mep} (a), we only show the switching between the hopfion and MAP with the same polarization, but it is also possible to switch between MAP states with opposite polarizations using a large field to saturating spins in the opposite direction.

Since the hopfion is topologically protected by the nonzero Hopf invariant, 
a topological transition must take place in the switching between the hopfion and MAP states.
To investigate this topological transition, 
we performed a minimal energy path (MEP) calculation between these two states~\cite{bessarab_method_2015, cortes_ortuno_thermal_2017, bessarab_annihilation_2017}.
The MEP calculation is carried out using the geodesic nudged elastic band (GNEB) method 
associated with the Hamiltonian in 
Eq. (\ref{eq:energy}). 
The stable spin textures from the energy minimizations are employed as the initial states in the MEP calculation.

%
Results from the MEP calculation are shown in Fig. \ref{fig:mep}(b).
There exists an energy barrier between the hopfion and the MAP state. 
Thus, an activation energy is required to enable the transition from the hopfion (MAP) to MAP (hopfion) state.
To capture details of the topological transition, 
we plot preimages of $\bf s=$(1,0,0) and $\bf s=$(-1,0,0) at the initial hopfion state, 
the barrier peak, the intermediate MEP state and the final MEP state (Fig.~\ref{fig:mep}(c)-(f)).
Transitioning from the hopfion state in (c) to the intermediate state (e),
the two linked preimages break and 
reconnect generating the monopole--antimonopole pair with a $2\pi$ rotation.
The two preimages are then topologically equivalent to those of the MAP state in Fig. \ref{fig:mep}(f), 
although they are twisted by $2\pi$. 
Relaxing from point (e) to to the MAP state of point (f), 
the preimages untwist to $\pi$, while the monopole and antimonopole move towards the top and bottom 
surface, respectively.
Videos of the transition also capture the transformation from a torus ($g=1$) to a trivial surface ($g=0$) for the preimages of $s_z=0$ (see movies in the Supplemental Materials). 
To create a hopfion from a MAP state, the reverse process is applied.
The preimages first rotate from $\pi$ to $2\pi$. 
The monopole--antimonopole pair move towards each other until they eliminate each other. 
Then each preimage becomes close-looped and linked with the other preimage.
To conclude, a $Q_H=1$ hopfion can be stabilized in a chiral magnet nanodisk 
sandwiched by two magnetic layers with PMA at zero external magnetic fields.
The hopfion is identified by its preimages and the Hopf invariant.
A MAP state is also stabilized at zero field in the proposed structure.
The hopfion (MAP) can be switched into a MAP (hopfion) state by applying a magnetic field.
The minimal energy path calculation reveals the topological transition between the hopfion and the MAP state.
%
%
3D magnetic imaging techniques such as the X-ray vector nanotomography 
could be a powerful tool for visualizing the spin texture of hopfion in real space~\cite{3D_imaging}.
The hopfion may exhibit fascinating electronic transport and dynamical properties due to its novel topology.
This work paves a way in the development of 3D spintronics 
and high dimensional memory architectures~\cite{3D_spintronics}.

\noindent
{\em Acknowledgements:} 
JZ acknowledges stimulating discussions with Jeffrey Teo. 
Conception and analytical works in this work were supported by the U.S. Department of Energy
(DOE), Office of Science, Basic Energy Sciences (BES) under Award
No. DE-SC0016424. 
Numerical simulations and part of the analytical work 
were supported as part of the Spins and Heat in Nanoscale Electronic Systems (SHINES) an 
Energy Frontier Research Center funded by the U.S. Department of Energy, Office of Science, 
Basic Energy Sciences under Award No. DE-SC0012670. 
Numerical simulations and collaborative travel between UCR and UNH were also supported by the NSF ECCS-1408168.


%

\end{document}